\documentclass[epjc3,twocolumn]{svjour3}

\usepackage{hep-paper}

\bibliography{bibliography}

\acronym{CHF}{Swiss franc}
\acronym{SM}{Standard Model}
\acronym{LLP}{long-lived particle}
\acronym{DM}{Dark Matter}
\acronym{BBN}{big bang nucleosynthesis}
\acronym{RPC}{resistive plate chamber}
\acronym{CMB}{cosmic microwave background}
\acronym[$\nu$MSM]{nuMSM}{Neutrino Minimal Standard Model}
\acronym{HNL}{\emph{heavy neutral lepton}}
\acronym{HECATE}{\emph{HErmetic CAvern TrackEr}}
\acronym{HADES}{\emph{HAdron-collider-cavern DEtector System}}
\acronym{THUNDERDOME}{\emph{Totally Hyper-UNrealistic DEtectoR in a huge DOME}}
\acronym{IP}{interaction point}

\shortacronym{FCC}{Future Circular Collider}
\shortacronym[\mbox{FCC-ee}]{FCCee}{electron FCC}
\shortacronym[\mbox{FCC-hh}]{FCChh}{hadron FCC}
\shortacronym{SPPC}{Super Proton-Proton Collider}
\shortacronym{CEPC}{Circular Electron Positron Collider}
\shortacronym{LHC}{Large Hadron Collider}
\shortacronym{ILC}{International Linear Collider}
\shortacronym{MATHUSLA}{MAssive Timing Hodoscope for Ultra-Stable neutraL PArticles}

\journalname{Eur.\ Phys.\ J.\ C}

\begin{document}

\title{HECATE}

\subtitle{A long-lived particle detector concept for the FCC-ee or CEPC}

\author{Marcin Chrząszcz\thanksref{marcin,krakow}
\and Marco Drewes\thanksref{marco,louvain}
\and Jan Hajer\thanksref{jan,louvain}}

\institute{
Institute of Nuclear Physics, Polish Academy of Sciences, ul.\ Radzikowskiego 152, 31--342 Kraków, Poland \label{krakow}
\and
Centre for Cosmology, Particle Physics and Phenomenology, Université catholique de Louvain, Louvain-la-Neuve B-1348, Belgium \label{louvain}
}

\date{November 2020}

\thankstext{marcin}{\email{marcin.chrzaszcz@ifj.edu.pl}}
\thankstext{marco}{\email{marco.drewes@uclouvain.be}}
\thankstext{jan}{\email{jan.hajer@uclouvain.be}}

\maketitle

\begin{abstract}
The next generation of circular high energy collider is expected to be a lepton collider, FCC-ee at CERN or CEPC in China.
However, the civil engineering concepts foresee to equip these colliders with bigger detector caverns than one would need for a lepton collider, so that they can be used for a hadron collider that may be installed in the same tunnel without further civil engineering.
This opens up the possibility to install extra instrumentation at the cavern walls to search for new long-lived particles at the lepton collider.
We use the example of heavy neutral leptons to show that such an installation could improve the sensitivity to the squared mixing parameter by almost half an order of magnitude.
\end{abstract}

\section{Introduction}

Future lepton colliders such as the \FCCee \cite{Abada:2019zxq} or \CEPC \cite{CEPCStudyGroup:2018rmc} have an extremely rich physics program \cite{Abada:2019lih, CEPCStudyGroup:2018ghi}.
In particular, they are outstanding intensity frontier machines that can not only study the properties of the electroweak and Higgs sector at unprecedented accuracy, but they can also search for feebly coupled hidden particles that have escaped detection at the \LHC due to their low production cross section.
Due to their small interaction strength hidden particles can have comparably long lifetimes.
Such new \LLPs appear in many extensions of the \SM of particle physics that can address open questions in particle physics and cosmology, such as the \DM, neutrino masses or baryogenesis, \cf \eg \cite{Curtin:2018mvb}.
In recent years many studies have investigated the sensitivity of the \LHC \cite{Alimena:2019zri} and other experiments \cite{Beacham:2019nyx} to \LLPs.
The clean environment of a lepton collider would offer even better perspectives for such searches \cite{Abada:2019zxq, CEPCStudyGroup:2018ghi}.

\LLPs are typically searched for at colliders through displaced signatures \cite{Alimena:2019zri}.
One constraint in this context is the volume of the main detectors, which limits the potential to search for particles with very long lifetimes.
For the \LHC, several dedicated detectors have been proposed to extend the reach to larger lifetimes, including FASER \cite{Feng:2017uoz}, \MATHUSLA \cite{Chou:2016lxi}, CODEX-b \cite{Gligorov:2017nwh}, Al3X \cite{Gligorov:2018vkc}, MAPP \cite{Mitsou:2020hmt}, and ANUBIS \cite{Bauer:2019vqk}.
While \MATHUSLA would be placed at the surface, the other proposals take advantage of existing cavities that can be instrumented.
At a future lepton collider the detectors will be smaller than those of the \LHC, which limits the sensitivity to long lifetimes.
However, the current planning for the \FCCee and \CEPC foresees to build a hadron collider in the same tunnel, namely the \FCChh or \SPPC \cite{Benedikt:2018csr, CEPC-SPPCStudyGroup:2015csa}.
For this reason, it has been proposed and is widely accepted that the detector caverns for the \FCCee will be much bigger than needed for a lepton collider, so that the \FCChh and its detectors can be installed in the same tunnel without major civil engineering effort.
In this Letter we point out that instrumenting this extra space could considerably increase the sensitivity of the \FCCee (or likewise the \CEPC) to \LLPs at a cost of only a few million \CHF.
We therefore propose to include such a \HECATE in future \FCCee and \CEPC studies.
\footnote{
In a first version of the manuscript the detector concept was called $\HADES$.
The names has been changed to \HECATE to avoid confusion with the existing experiment at GSI.
}

\section{Possible implementation and cost estimate}

A possible implementation of the \HECATE detector would consist of \RPCs or scintillator plates, constructed from extruded scintillating bars, located around the cavern walls and forming a $4\pi$ detector.
Such an hermetic detector design maximizes the fiducial volume and allows to discriminate against events originating from outside the detector cavern.
The inner detector and muon chambers act as a veto able to reject \SM particles from the primary vertex.
In order to obtain timing information and to distinguish particles from cosmic background, the \HECATE detector should have at least two layers of detector material separated by a sizable distance.
For reliable tracking, at least four layers, along with a smaller size and/or optimised geometry of the detector plates, would be required.
The biggest challenge of such detector is the control of the background that will originate from two sources: cosmics and neutrinos produced at collision point.
The first can be handled using timing information, which permits to distinguish particles coming from within the cavern from ones coming from outside the cavern.
We expect neutrinos that are produced from cosmic rays and decay within the cavern volume to be more problematic.
However, this background requires a detailed Monte Carlo simulation, which lies beyond the scope of this Letter.
On the other hand, the feasibility of such a rejection is studied for the \MATHUSLA detector, see \cite{Alpigiani:2020tva} and references therein.
We expect a lower background for \HECATE, as \MATHUSLA would be exposed to more cosmics due to its location on the surface.
More dangerous is the background from \SM neutrinos produced in process such as $e^+ e^- \mathpunct\to Z \mathpunct\to \nu \overline \nu$, as neutrinos may interact with the detector material resulting in detectable charged particles.
These type of events can be rejected in different ways, depending on where the interaction takes place.
Events originating from the beam pipe can be rejected by using the tracker as a veto and interactions within the inner detectors can be rejected using its outer layers.
Additionally, in the case that the background remains too large, one could install an additional tracker layer outside the detector as a veto system, that would be able to reject \SM neutrino interactions in the outer layers of the usual detector.
The present Letter is a simple proof-of-principle, and we postpone a detailed investigation of such backgrounds to future work.
We note in passing that the \MATHUSLA collaboration, which faces similar issues, has concluded that they are under control, see \cite{Alpigiani:2020tva} and references therein.

For a horizontally cylindrical cavern with a radius of \unit[15]{m} and length of \unit[50]{m}, plates of $\unit[1]{m^2}$ surface would provide $\mathpunct \sim 6000$ readout channels for the scintillating bars.
The main cost of such detector would then be the cost of the scintillators.
Assuming a thickness of \unit[1]{cm} for a single panel the cost would amount to \unit[3--5]{M\CHF} with current prices.
This assumes that the used scintillator is EJ-200, which has a long optical attenuation length and fast timing.
The cost could be significantly reduced if a cheaper alternative is used that still matches the specifications required for such a detector design.
The cost of the readout electronics can be estimated based on the \mbox{Sci-Fi} detector from LHCb \cite{LHCbCollaboration:2014tuj}.
On this basis, the readout electronics together with the clear and wave-shifting fibers needed for the scintillator would cost around \unit[30]{\CHF} per channel.
Hence, the total cost of the detector would be below \unit[5]{M\CHF} per layer.
This estimate assumes present day technology and one can expect that better technology can be purchased at lower prices at the time the \FCCee will be built.

\section{Sensitivity estimate}

We estimate the \HECATE sensitivity for right-handed neutrinos, a type of \HNLs.
The existence of these \HNLs is predicted by well-motivated extensions of the \SM, in particular the type-I seesaw mechanism \cite{Minkowski:1977sc, GellMann:1980vs, Mohapatra:1979ia, Yanagida:1980xy, Schechter:1980gr, Schechter:1981cv}, leptogenesis \cite{Fukugita:1986hr}, and as \DM candidates \cite{Dodelson:1993je}.
They have been a benchmark scenario for \FCCee sensitivity estimates from early stages \cite{Blondel:2014bra}.
The properties of \HNLs are characterised by their Majorana mass $M$ and the mixing angles $\theta_a$ that determine the suppression of their weak interactions relative to that of ordinary neutrinos.
The implications of the heavy neutrinos' existence strongly depend on the magnitude of $M$, see \eg \cite{Drewes:2013gca} for a review.
\footnote{For further details on specific aspects we refer the reader to reviews on leptogenesis \cite{Garbrecht:2018mrp, Bodeker:2020ghk} and the perspectives to test it \cite{Chun:2017spz}, sterile neutrino \DM \cite{Adhikari:2016bei, Boyarsky:2018tvu} and experimental searches for heavy neutrinos \cite{Atre:2009rg, Deppisch:2015qwa, Cai:2017mow, Antusch:2016ejd}.}
The seesaw mechanism requires at least two \HNLs to explain the light neutrino oscillation data.
However, for the purpose of the present study a simplified model with a single \HNL denoted by $N$ suffices,
\begin{multline} \label{eq:weak intraction}
 \mathcal L \supset
- \frac{m_W}{v} \overline N \theta^*_a \gamma^\mu e_{La} W^+_\mu
- \frac{m_Z}{\sqrt 2 v} \overline N \theta^*_a \gamma^\mu \nu_{La} Z_\mu \\
- \frac{M}{v} \theta_a h \overline{\nu_L}_a N
+ \text{h.c.} \ ,
\end{multline}
where $e_{La}$ and $\nu_{La}$ are the charged and neutral \SM leptons, respectively, $Z$ and $W$ are the weak gauge bosons with masses $m_Z$ and $m_W$, and $h$ is the physical Higgs field after spontaneous breaking of the electroweak symmetry by the expectation value $v$.

At a lepton collider the \HNLs are primarily produced from the decay of on-shell $Z$-bosons.
The \HNL production cross section in the process $Z\to \nu N$ can be estimated as \cite{Blondel:2014bra}
\footnote{The sub-dominant $N$ production in the decay of $B$-mesons generated in the process $Z \mathpunct \to b \bar b$ has \eg been studied in \cite{Chun:2019nwi}.}
\begin{equation}
\sigma_N \simeq 2\sigma_Z \operatorname{BR}(Z\mathpunct\to\nu\overline\nu) U^2 \left(1 - \frac{M^2}{m_Z^2}\right)^{\!2} \left(1 + \frac{M^2}{m_Z^2}\right) \ ,
\end{equation}
their decay rate is roughly $\Gamma_N \simeq 12 U^2 M^5 G_F^2 /(96\pi^3)$ with $G_F$ the Fermi constant and $U^2=\sum U_a^2$ with $U_a^2=\abs{\theta_a}^2$.
\footnote{The decay rates of \HNLs into \SM particles have been computed by many different authors \cite{Johnson:1997cj, Gorbunov:2007ak, Atre:2009rg, Canetti:2012kh, Bondarenko:2018ptm, Ballett:2019bgd, Coloma:2020lgy, Pascoli:2018heg, deVries:2020qns}, they overall more or less agree with each other.}
The number of events that can be observed in a spherical detector with an integrated luminosity $L$ can then be estimated as
\footnote{
In \cite{Drewes:2019vjy} we have confirmed in a proper simulation that the estimate \eqref{eq:observed events} works reasonably well even at the \LHC main detectors if one takes the average over the \HNL momentum distribution.
In the much cleaner environment of a lepton collider we expect that it works even better.
}
\begin{equation} \label{eq:observed events}
N_\text{obs} \simeq L \sigma_N \left[\exp(-\frac{l_0}{\lambda_N}) - \exp(-\frac{l_1}{\lambda_N}) \right] \ .
\end{equation}
Here $\lambda_N = \flatfrac{\beta \gamma}{\Gamma_N}$ is the \HNL decay length in the laboratory frame, $l_0$ and $l_1$ denote the minimal and maximal distance from the \IP where the detector can detect an \HNL decay into charged particles.
If the $Z$-boson decays at rest we can set $\beta\gamma=(m_Z^2 - M^2)/(2 m_Z M)$.
\footnote{We confirmed that this is a good approximation by generating the \HNL momentum distribution with \software[8.2]{Pythia} \cite{Sjostrand:2014zea} (including initial state radiation) and averaging \eqref{eq:observed events} over this distribution.}
We then replaced one of the neutrinos with the \HNL with a given mass.
We have considered masses spanning from \unit[1]{GeV} up to $m_Z$ in steps of \unit[1]{GeV}.
\footnote{Natural units with $c = 1$ are used throughout this Letter.}

\begin{figure}
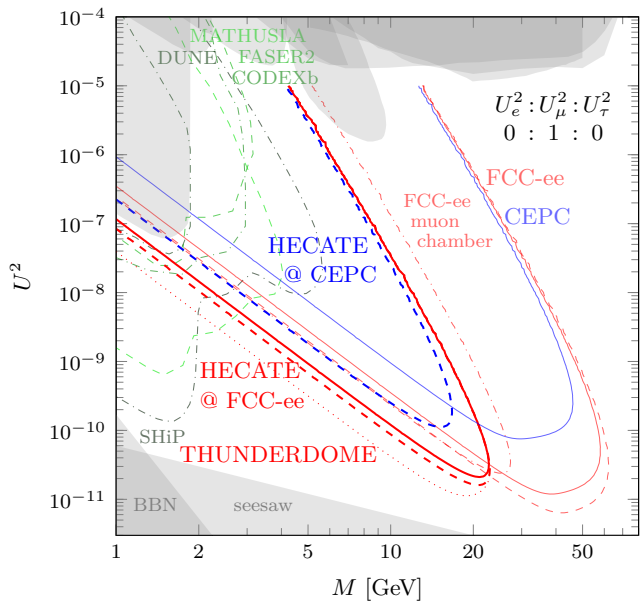

\graphic{U2mu-hecate}
\caption{
Comparison of the sensitivities for nine signal events that can be achieved at the \FCCee with $2.5\cdot10^{12}$ $Z$-bosons (red) or \CEPC with $3.5\cdot10^{11}$ $Z$-bosons (blue).
The faint solid curves show the main detector sensitivity ($l_0 = \unit[5]{mm}$, $l_1 = \unit[1.22]{m}$).
The faint dash-dotted curve indicates the additional gain if the muon chambers are used at the \FCCee ($l_0 = \unit[1.22]{m}$, $l_1 = \unit[4]{m}$).
The thick curves show the sensitivity of \HECATE with $l_0 = \unit[4]{m}$, $l_1 = \unit[15]{m}$ (solid) and $l_0 = \unit[4]{m}$, $l_1 = \unit[25]{m}$ (dashed), respectively.
Finally, the faint dashed red line shows the \FCCee main detector sensitivity with $5\cdot10^{12}$ $Z$-bosons, corresponding to the luminosity at two \IPs.
For comparison we indicate the expected sensitivity of selected other experiments with the different green curves as indicated in the plot \cite{Chou:2016lxi,Feng:2017uoz,Gligorov:2017nwh,SHiP:2018xqw,Ballett:2019bgd}.
The gray areas in the upper part of the plot show the region excluded by past experiments \cite{Bergsma:1985is, Sirunyan:2018mtv, Aad:2019kiz, Abreu:1996pa, Artamonov:2014urb, Aaij:2016xmb, Antusch:2017hhu, Vaitaitis:1999wq, Bernardi:1987ek}, the grey areas at the bottom mark the regions that are disfavoured by $\BBN$ and neutrino oscillation data in the $\nuMSM$ (\enquote{seesaw}).
} \label{fig:comparison}
\end{figure}

In \cref{fig:comparison} we show the expected gain in sensitivity that can be achieved with \HECATE (thick curves; red and blue encoding the \FCCee and \CEPC, respectively; solid and dashed, corresponding to $l_1 = \unit[15]{m}$ or $l_1 = \unit[25]{m}$) in comparison to using only the inner detector (faint red and blue curve for \FCCee and \CEPC) with $2.5\cdot10^{12}$ and $3.5\cdot10^{11}$ $Z$-bosons, respectively.
These numbers refer to the expected integrated luminosity during the $Z$-pole run at one \IP.
We display lines for nine detected signal events approximately corresponding to  the \unit[5]{\sigma} discovery region under the assumption of a single background event.
The actual sensitivity of \HECATE should lie somewhere between the two thick red curves, as the approximately cylindrical detector extents from the \IP \unit[15]{m} in radial direction and \unit[25]{m} in beam direction.
The improvement with \HECATE is almost half an order of magnitude in $U^2$ for given $M$.
This can be understood by recalling that the region on the lower left side of the sensitivity region corresponds to decay lengths that greatly exceed the detector size.
In this regime the exponentials in \eqref{eq:observed events} can be expanded in $l_1/\lambda$ and $l_0/\lambda$.
For $l_1 \gg l_0$ the number of events is simply given by
\begin{equation} \label{eq:long life limit}
N_\text{obs} \simeq L \sigma_N  \frac{\Gamma_N l_1}{\beta\gamma} \propto L  U^4 \frac{M^5 l_1}{\beta\gamma} \ .
\end{equation}
Hence, the value of $U^2$ that leads to a given number of events for fixed $M$ scales as $\mathpunct \propto 1/\sqrt{l_1}$.
For the inner detector we assume a radius of \unit[1.22]{m} within which displaced vertices can be detected.
This corresponds to the size of the ECAL designed for the \ILC \cite{Behnke:2013lya}, we use it as an estimate for the dimensions of the \FCCee or \CEPC detectors, which are to be determined.
\footnote{
In \cite{Antusch:2016vyf} the number \unit[2.49]{m} was used, corresponding to the dimensions of the \ILC HCAL.
The gain in sensitivity can be estimated with \eqref{eq:long life limit}, as outlined below, and lies somewhere between our main detector and muon chamber lines in \cref{fig:comparison}.
}
Hence, one can expect relative sensitivity gains $\mathpunct\propto \sqrt{\unit[1.22]{m}/l_1}$.
Relation \eqref{eq:long life limit} also permits to estimate the sensitivity gain by increasing the integrated luminosity $L$ (\eg by extending the $Z$-pole run or by considering more than one \IP):
Increasing $L$ by a given factor has the same effect as increasing $l_1$ by the same factor.
Hence, for given $M$, the value of $U^2$ needed to achieve a given number of events scales as $\mathpunct \propto 1/\sqrt L$.
As an example we display the gain in sensitivity that could be achieved with the inner detector by doubling the integrated luminosity of the $Z$-pole run (faint dashed red curve).
The scaling of all lines in the plot with increased integrated luminosity can be estimated by comparing the faint dashed red line to the faint solid red curve, and with relation \eqref{eq:long life limit}.
All \HECATE lines would scale with the number of $Z$-bosons in the same way as the main detector line, we omit them here to keep the plot readable.
We further do not show all lines for the \CEPC, the omitted ones would give similar relative sensitivity gains as for \FCCee.

Further, we estimate the potential sensitivity gain from performing a search in the muon chambers at the \FCCee (faint dash-dotted red curve).
This idea, originally proposed in \cite{Bobrovskyi:2011vx, Bobrovskyi:2012dc}, has been applied to \HNL searches at the \LHC in \cite{Boiarska:2019jcw, Drewes:2019fou}.
As suggested by the scaling above, the gain is considerably lower than what could be done with \HECATE.
Finally, one may wonder whether it is worth to dig even bigger caverns to host dedicated \LLP detectors.
The scaling $\mathpunct \propto 1/\sqrt{l_1}$ of the sensitivity to $U^2$ for given $M$ implies that the costs for civil engineering would quickly grow.
For illustrative purposes we  add the sensitivity that could be achieved with the very unrealistic \THUNDERDOME concept at \FCCee ($l_0 = \unit[4]{m}$, $l_1 = \unit[100]{m}$, dotted red line).
However, for other \LLP models with a different scaling the return of investment might be better.

It should be said that \cref{fig:comparison} is very conservative as far as the sensitivity gain with \HECATE relative to the inner detector is concerned because we have assumed \unit[100]{\%} efficiency and no backgrounds for both.
For \HECATE these assumptions are semi-realistic, as the inner detector can be used as a veto.
In contrast, in the inner detector the reconstruction efficiency for displaced vertices rapidly decreases as a function of displacement.
This dependence has been studied for the \LHC \cite{Aaboud:2017iio, ATLAS:2019wqx, Aad:2019kiz} and LEP \cite{Buskulic:1994wz, Abreu:1996pa}, but a realistic estimate for the \FCCee detectors would require detailed simulations.
In the present note, which is a proof-of-principle, we therefore choose to make the same assumptions for \HECATE and the inner detector and therefore underestimate the relative sensitivity gain that could be achieved with \HECATE.

It is instructive to compare the \HECATE sensitivity to existing constraints and to the reach of other upcoming or proposed experiments.
An updated summary of relevant experimental constraints can be found in \cite{Chrzaszcz:2019inj}.
\footnote{
The sensitivity of some experiments depend on the number $n$ of \HNL flavours, the mass of the lightest \SM neutrino $m_\text{lightest}$, and on the flavour mixing pattern, \ie the relative size of the mixings $\abs{\theta_a}^2$ with individual \SM generations.
It is therefore difficult to make an apple-to-apple comparison.
The sensitivity of direct searches for displaced searches at accelerators in good approximation only depends on the flavour mixing pattern.
In contrast to that, searches that rely on lepton number violating signatures strongly depend on the \HNL mass spectrum and light neutrino properties \cite{Drewes:2019byd}.
}
In \cref{fig:comparison} we display the exclusion region of several experiments under the assumption that the \HNL exclusively mix with the second \SM generation ($U^2=\abs{\theta_\mu}^2$) \cite{Bergsma:1985is, Sirunyan:2018mtv, Aad:2019kiz, Abreu:1996pa, Artamonov:2014urb, Aaij:2016xmb, Antusch:2017hhu, Vaitaitis:1999wq, Bernardi:1987ek}.
Indirect searches, on the other hand, strongly depend on the properties of light neutrinos, \cf \eg \cite{Chrzaszcz:2019inj, Drewes:2016jae, Fernandez-Martinez:2016lgt, Hernandez:2016kel, Drewes:2015iva, Gorbunov:2014ypa, Antusch:2014woa} for a detailed discussion.
In particular, lower bounds on the individual $\abs{\theta_a}$ from neutrino oscillation data can only be imposed under specific model assumptions, and the lower bound on their sum $U^2$ scales as $m_\text{lightest}/M$.
As an indicator, we add the corresponding lower \enquote{seesaw} bound in the \nuMSM \cite{Asaka:2005an, Asaka:2005pn} as a gray area, assuming normal ordering of the light neutrinos.
There is also a bound on the lifetime of the $N$ from the requirement to decay before \BBN in the early universe, which again depends on the flavour mixing pattern.
We here display the bound from \cite{Boyarsky:2020dzc} under the assumption $U^2=\abs{\theta_\mu}^2$.
\footnote{
In the recent works \cite{Sabti:2020yrt, Boyarsky:2020dzc} it was assumed that the \HNLs are in thermal equilibrium in the early universe, which is in general not true for small $U^2$.
In \cite{Domcke:2020ety} it was, however, pointed out that smaller mixing angles are ruled out by the cosmological history between \BBN and the \CMB decoupling \cite{Vincent:2014rja}, and various other effects of \HNL decays (dissociation of nuclei \cite{Domcke:2020ety}, effect the \CMB anisotropies \cite{Poulin:2016anj} heating up the intergalactic medium \cite{Diamanti:2013bia}).
In the mass range considered here, the bounds from \cite{Sabti:2020yrt, Boyarsky:2020dzc} can therefore be regarded as \enquote{hard} unless one considers $U^2$ that are so tiny that the \HNL are never produced in significant quantities.
}

To put \HECATE into the context of the future experimental program in particle physics, we indicate the sensitivity of selected other proposed or planned experiments, as indicated in the plot \cite{Chou:2016lxi,Feng:2017uoz,Gligorov:2017nwh,SHiP:2018xqw,Ballett:2019bgd}.
\HECATE would be complementary to other proposals and help to fill the sensitivity gap between the reach of future lepton colliders limited by the volume of their detectors and fixed target experiment limited by the $D$-meson threshold, \cf \cref{fig:comparison}.

Finally, one may compare the reach to the parameter region where leptogenesis is possible in well-motivated scenarios.
The most updated parameter space scans for the \nuMSM (practically $n=2$) and the model with $n=3$ can be found in \cite{Klaric:2020lov} and \cite{Abada:2018oly}, respectively.
In both cases \HECATE can probe regions deep inside the leptogenesis parameter space.
In the future, it would be interesting to study how well \HECATE could measure the \HNL properties, so that, in case any \HNLs are discovered, one could address the question whether or not these particles are indeed responsible for the origin of matter in the universe.
For the \FCC inner detector this has been done in \cite{Caputo:2016ojx, Antusch:2017pkq}.

\section{Discussion and conclusions}

In this Letter we point out that an instrumentation of the large detector caverns that are planned at future circular lepton colliders could considerably increase the sensitivity of searches for \LLPs.
The main difference between the \HECATE concept and other proposals (such as surface detectors \cite{Chou:2016lxi,Wang:2019xvx}, forward detectors \cite{Chou:2016lxi} or installations on the cavern ceiling \cite{Wang:2019xvx}) lies in the approximate $4\pi$ solid angle coverage that can be achieved by covering the cavern walls with detectors.
The possibility to install a far detector that is sensitive to displacements over \unit[20]{m} with such large angular coverage without extra civil engineering is a unique opportunity resulting from the fact that the caverns for \FCCee or \CEPC are expected to be designed to host a hadron collider.
In addition to the large caverns, there will also be large vertical shafts over the chambers for the detector installation ($\unit[16]{m}$ diameter planned for the \CEPC, and over \unit[20]{m} for the \FCC at at least two \IPs), which could be used to install further instrumentation (as proposed for the \LHC with ANUBIS).

The proposed solution based on scintillators should be regarded as an example and proof-of-principle.
There are many other detectors that will reach timing and efficiency requirements, such as \RPCs.
The final choice would be based on detailed study of balance between the cost and the detector performance.

Using this example, we estimate that a detector of the \HECATE type could increase the \FCC sensitivity to heavy neutrinos (or \HNLs) with masses below $m_Z$ by almost half an order of magnitude at a cost of the order of ten million \CHF, conservatively assuming present detector technology and prices.
This improvement would help to fill the sensitivity gap for \HNL masses between the $B$-meson mass (below which fixed target experiments are highly sensitive) and the regime that can be covered by the \FCCee or \CEPC main detectors.
Heavy neutrinos are only one example for \LLPs, which we have chosen for illustrative purposes here.
\HECATE could open many other portals to dark sectors that can potentially address some of the \enquote{big questions} in particle physics and cosmology, including \DM, baryogenesis, cosmic inflation and neutrino masses.

\subsection*{Acknowledgments}

The authors would like to thank Alain Blondel, David Curtin, Albert De Roeck, Rebeca Gonzalez Suarez, Elena Graverini, and Manqi Ruan for useful comments on this Letter.
The work of M.C.\ is funded by the Polish National Agency for Academic Exchange under the Bekker program.
M.C.\ is also grateful for the funding from the European Union's Horizon 2020 research and innovation programme under grant \no{951754}.

\printbibliography

@article{SHiP:2018xqw,
    author = "Ahdida, C. and others",
    collaboration = "SHiP",
    title = "{Sensitivity of the SHiP experiment to Heavy Neutral Leptons}",
    eprint = "1811.00930",
    archivePrefix = "arXiv",
    primaryClass = "hep-ph",
    doi = "10.1007/JHEP04(2019)077",
    journal = "JHEP",
    volume = "04",
    pages = "077",
    year = "2019"
}

@article{Abada:2019zxq,
    author = "Abada, A. and others",
    collaboration = "FCC",
    title = "{FCC-ee: The Lepton Collider}: {Future Circular Collider Conceptual Design Report Volume 2}",
    reportNumber = "CERN-ACC-2018-0057",
    doi = "10.1140/epjst/e2019-900045-4",
    journal = "Eur. Phys. J. ST",
    volume = "228",
    number = "2",
    pages = "261--623",
    year = "2019"
}

@article{Antusch:2016vyf,
      author         = "Antusch, Stefan and Cazzato, Eros and Fischer, Oliver",
      title          = "{Displaced vertex searches for sterile neutrinos at
                        future lepton colliders}",
      journal        = "JHEP",
      volume         = "12",
      year           = "2016",
      pages          = "007",
      doi            = "10.1007/JHEP12(2016)007",
      eprint         = "1604.02420",
      archivePrefix  = "arXiv",
      primaryClass   = "hep-ph",
      SLACcitation   = "%%CITATION = ARXIV:1604.02420;%%"
}

@article{Chun:2019nwi,
      author         = "Chun, Eung Jin and Das, Arindam and Mandal, Sanjoy and
                        Mitra, Manimala and Sinha, Nita",
      title          = "{Sensitivity of Lepton Number Violating Meson Decays in
                        Different Experiments}",
      journal        = "Phys. Rev.",
      volume         = "D100",
      year           = "2019",
      number         = "9",
      pages          = "095022",
      doi            = "10.1103/PhysRevD.100.095022",
      eprint         = "1908.09562",
      archivePrefix  = "arXiv",
      primaryClass   = "hep-ph",
      reportNumber   = "OU-HEP-1016, IP/BBSR/2019-4",
      SLACcitation   = "%%CITATION = ARXIV:1908.09562;%%"
}

@article{Behnke:2013lya,
      author         = "Abramowicz, Halina and others",
      editor         = "Behnke, Ties and Brau, James E. and Burrows, Philip N.
                        and Fuster, Juan and Peskin, Michael and Stanitzki, Marcel
                        and Sugimoto, Yasuhiro and Yamada, Sakue and Yamamoto,
                        Hitoshi",
      title          = "{The International Linear Collider Technical Design
                        Report - Volume 4: Detectors}",
      year           = "2013",
      eprint         = "1306.6329",
      archivePrefix  = "arXiv",
      primaryClass   = "physics.ins-det",
      reportNumber   = "ILC-REPORT-2013-040, ANL-HEP-TR-13-20,
                        BNL-100603-2013-IR, IRFU-13-59, CERN-ATS-2013-037,
                        COCKCROFT-13-10, CLNS-13-2085, DESY-13-062,
                        FERMILAB-TM-2554, IHEP-AC-ILC-2013-001, INFN-13-04-LNF,
                        JAI-2013-001, JINR-E9-2013-35, JLAB-R-2013-01,
                        KEK-REPORT-2013-1, KNU-CHEP-ILC-2013-1, LLNL-TR-635539,
                        SLAC-R-1004, ILC-HIGRADE-REPORT-2013-003",
      SLACcitation   = "%%CITATION = ARXIV:1306.6329;%%"
}

@article{Wang:2019xvx,
      author         = "Wang, Zeren Simon and Wang, Kechen",
      title          = "{Physics with far detectors at future lepton colliders}",
      journal        = "Phys. Rev.",
      volume         = "D101",
      year           = "2020",
      number         = "7",
      pages          = "075046",
      doi            = "10.1103/PhysRevD.101.075046",
      eprint         = "1911.06576",
      archivePrefix  = "arXiv",
      primaryClass   = "hep-ph",
      reportNumber   = "APCTP Pre2019-024",
      SLACcitation   = "%%CITATION = ARXIV:1911.06576;%%"
}

@article{Benedikt:2018csr,
    author = "Abada, A. and others",
    collaboration = "FCC",
    title = "{FCC-hh: The Hadron Collider}: {Future Circular Collider Conceptual Design Report Volume 3}",
    reportNumber = "CERN-ACC-2018-0058",
    doi = "10.1140/epjst/e2019-900087-0",
    journal = "Eur. Phys. J. ST",
    volume = "228",
    number = "4",
    pages = "755--1107",
    year = "2019"
}

@article{CEPC-SPPCStudyGroup:2015csa,
    author = "Ahmad, Muhammd and others",
    title = "{CEPC-SPPC Preliminary Conceptual Design Report. 1. Physics and Detector}",
    reportNumber = "IHEP-CEPC-DR-2015-01, IHEP-TH-2015-01, IHEP-EP-2015-01",
    month = "3",
    year = "2015"
}

@article{Mitsou:2020hmt,
      author         = "Mitsou, Vasiliki A.",
      title          = "{MoEDAL physics results and future plans}",
      booktitle      = "{Proceedings, 19th Hellenic School and Workshops on
                        Elementary Particle Physics and Gravity (CORFU2019):
                        Corfu, Greece, August 31 - September 25, 2020}",
      collaboration  = "MoEDAL",
      journal        = "PoS",
      volume         = "CORFU2019",
      year           = "2020",
      pages          = "009",
      doi            = "10.22323/1.376.0009",
      SLACcitation   = "%%CITATION = POSCI,CORFU2019,009;%%"
}

@article{Antusch:2014woa,
      author         = "Antusch, Stefan and Fischer, Oliver",
      title          = "{Non-unitarity of the leptonic mixing matrix: Present
                        bounds and future sensitivities}",
      journal        = "JHEP",
      volume         = "10",
      year           = "2014",
      pages          = "094",
      doi            = "10.1007/JHEP10(2014)094",
      eprint         = "1407.6607",
      archivePrefix  = "arXiv",
      primaryClass   = "hep-ph",
      reportNumber   = "MPP-2014-313",
      SLACcitation   = "%%CITATION = ARXIV:1407.6607;%%"
}

@article{Fernandez-Martinez:2016lgt,
      author         = "Fernandez-Martinez, Enrique and Hernandez-Garcia, Josu
                        and Lopez-Pavon, Jacobo",
      title          = "{Global constraints on heavy neutrino mixing}",
      journal        = "JHEP",
      volume         = "08",
      year           = "2016",
      pages          = "033",
      doi            = "10.1007/JHEP08(2016)033",
      eprint         = "1605.08774",
      archivePrefix  = "arXiv",
      primaryClass   = "hep-ph",
      SLACcitation   = "%%CITATION = ARXIV:1605.08774;%%"
}

@article{Drewes:2015iva,
      author         = "Drewes, Marco and Garbrecht, Björn",
      title          = "{Combining experimental and cosmological constraints on
                        heavy neutrinos}",
      journal        = "Nucl. Phys.",
      volume         = "B921",
      year           = "2017",
      pages          = "250-315",
      doi            = "10.1016/j.nuclphysb.2017.05.001",
      eprint         = "1502.00477",
      archivePrefix  = "arXiv",
      primaryClass   = "hep-ph",
      reportNumber   = "TUM-HEP-979-15",
      SLACcitation   = "%%CITATION = ARXIV:1502.00477;%%"
}

@article{Hernandez:2016kel,
      author         = "Hernández, P. and Kekic, M. and López-Pavón, J. and
                        Racker, J. and Salvado, J.",
      title          = "{Testable Baryogenesis in Seesaw Models}",
      journal        = "JHEP",
      volume         = "08",
      year           = "2016",
      pages          = "157",
      doi            = "10.1007/JHEP08(2016)157",
      eprint         = "1606.06719",
      archivePrefix  = "arXiv",
      primaryClass   = "hep-ph",
      SLACcitation   = "%%CITATION = ARXIV:1606.06719;%%"
}

@article{Gorbunov:2014ypa,
      author         = "Gorbunov, Dmitry and Timiryasov, Inar",
      title          = "{Testing $\nu$MSM with indirect searches}",
      journal        = "Phys. Lett.",
      volume         = "B745",
      year           = "2015",
      pages          = "29-34",
      doi            = "10.1016/j.physletb.2015.02.060",
      eprint         = "1412.7751",
      archivePrefix  = "arXiv",
      primaryClass   = "hep-ph",
      reportNumber   = "INR-TH-2014-035",
      SLACcitation   = "%%CITATION = ARXIV:1412.7751;%%"
}

@article{Drewes:2019byd,
    author = "Drewes, Marco and Klari\'c, Juraj and Klose, Philipp",
    title = "{On Lepton Number Violation in Heavy Neutrino Decays at Colliders}",
    eprint = "1907.13034",
    archivePrefix = "arXiv",
    primaryClass = "hep-ph",
    doi = "10.1007/JHEP11(2019)032",
    journal = "JHEP",
    volume = "19",
    pages = "032",
    year = "2020"
}

@article{Asaka:2005an,
      author         = "Asaka, Takehiko and Blanchet, Steve and Shaposhnikov,
                        Mikhail",
      title          = "{The $\nu$MSM, dark matter and neutrino masses}",
      journal        = "Phys. Lett.",
      volume         = "B631",
      year           = "2005",
      pages          = "151-156",
      doi            = "10.1016/j.physletb.2005.09.070",
      eprint         = "hep-ph/0503065",
      archivePrefix  = "arXiv",
      primaryClass   = "hep-ph",
      SLACcitation   = "%%CITATION = HEP-PH/0503065;%%"
}

@article{Asaka:2005pn,
      author         = "Asaka, Takehiko and Shaposhnikov, Mikhail",
      title          = "{The $\nu$MSM, dark matter and baryon asymmetry of the
                        universe}",
      journal        = "Phys. Lett.",
      volume         = "B620",
      year           = "2005",
      pages          = "17-26",
      doi            = "10.1016/j.physletb.2005.06.020",
      eprint         = "hep-ph/0505013",
      archivePrefix  = "arXiv",
      primaryClass   = "hep-ph",
      SLACcitation   = "%%CITATION = HEP-PH/0505013;%%"
}

@article{CEPCStudyGroup:2018rmc,
    collaboration = "CEPC Study Group",
    title = "{CEPC Conceptual Design Report: Volume 1 --- Accelerator}",
    eprint = "1809.00285",
    archivePrefix = "arXiv",
    primaryClass = "physics.acc-ph",
    reportNumber = "IHEP-CEPC-DR-2018-01, IHEP-AC-2018-01",
    month = "9",
    year = "2018"
}

@article{Abada:2019lih,
    author = "Abada, A. and others",
    collaboration = "FCC",
    title = "{FCC Physics Opportunities}: {Future Circular Collider Conceptual Design Report Volume 1}",
    reportNumber = "CERN-ACC-2018-0056",
    doi = "10.1140/epjc/s10052-019-6904-3",
    journal = "Eur. Phys. J. C",
    volume = "79",
    number = "6",
    pages = "474",
    year = "2019"
}

@article{CEPCStudyGroup:2018ghi,
    author = "Dong, Mingyi and others",
    editor = "Guimar\~aes da Costa, Jo\~ao Barreiro and others",
    collaboration = "CEPC Study Group",
    title = "{CEPC Conceptual Design Report: Volume 2 --- Physics \& Detector}",
    eprint = "1811.10545",
    archivePrefix = "arXiv",
    primaryClass = "hep-ex",
    reportNumber = "IHEP-CEPC-DR-2018-02, IHEP-EP-2018-01, IHEP-TH-2018-01",
    month = "11",
    year = "2018"
}

@article{Curtin:2018mvb,
    author = "Curtin, David and others",
    title = "{Long-Lived Particles at the Energy Frontier: The MATHUSLA Physics Case}",
    eprint = "1806.07396",
    archivePrefix = "arXiv",
    primaryClass = "hep-ph",
    reportNumber = "FERMILAB-PUB-18-264-T",
    doi = "10.1088/1361-6633/ab28d6",
    journal = "Rept. Prog. Phys.",
    volume = "82",
    number = "11",
    pages = "116201",
    year = "2019"
}

@article{Alimena:2019zri,
    author = "Alimena, Juliette and others",
    title = "{Searching for Long-Lived Particles beyond the Standard Model at the Large Hadron Collider}",
    eprint = "1903.04497",
    archivePrefix = "arXiv",
    primaryClass = "hep-ex",
    doi = "10.1088/1361-6471/ab4574",
    journal = "J. Phys. G",
    volume = "47",
    number = "9",
    pages = "090501",
    year = "2020"
}

@article{Beacham:2019nyx,
    author = "Beacham, J. and others",
    title = "{Physics Beyond Colliders at CERN: Beyond the Standard Model Working Group Report}",
    eprint = "1901.09966",
    archivePrefix = "arXiv",
    primaryClass = "hep-ex",
    reportNumber = "CERN-PBC-REPORT-2018-007",
    doi = "10.1088/1361-6471/ab4cd2",
    journal = "J. Phys. G",
    volume = "47",
    number = "1",
    pages = "010501",
    year = "2020"
}

@article{Feng:2017uoz,
    author = "Feng, Jonathan L. and Galon, Iftah and Kling, Felix and Trojanowski, Sebastian",
    title = "{ForwArd Search ExpeRiment at the LHC}",
    eprint = "1708.09389",
    archivePrefix = "arXiv",
    primaryClass = "hep-ph",
    reportNumber = "UCI-TR-2017-08",
    doi = "10.1103/PhysRevD.97.035001",
    journal = "Phys. Rev. D",
    volume = "97",
    number = "3",
    pages = "035001",
    year = "2018"
}

@article{Chou:2016lxi,
    author = "Chou, John Paul and Curtin, David and Lubatti, H.J.",
    title = "{New Detectors to Explore the Lifetime Frontier}",
    eprint = "1606.06298",
    archivePrefix = "arXiv",
    primaryClass = "hep-ph",
    doi = "10.1016/j.physletb.2017.01.043",
    journal = "Phys. Lett. B",
    volume = "767",
    pages = "29--36",
    year = "2017"
}

@article{Gligorov:2017nwh,
    author = "Gligorov, Vladimir V. and Knapen, Simon and Papucci, Michele and Robinson, Dean J.",
    title = "{Searching for Long-lived Particles: A Compact Detector for Exotics at LHCb}",
    eprint = "1708.09395",
    archivePrefix = "arXiv",
    primaryClass = "hep-ph",
    doi = "10.1103/PhysRevD.97.015023",
    journal = "Phys. Rev. D",
    volume = "97",
    number = "1",
    pages = "015023",
    year = "2018"
}

@article{Gligorov:2018vkc,
    author = "Gligorov, Vladimir V. and Knapen, Simon and Nachman, Benjamin and Papucci, Michele and Robinson, Dean J.",
    title = "{Leveraging the ALICE/L3 cavern for long-lived particle searches}",
    eprint = "1810.03636",
    archivePrefix = "arXiv",
    primaryClass = "hep-ph",
    doi = "10.1103/PhysRevD.99.015023",
    journal = "Phys. Rev. D",
    volume = "99",
    number = "1",
    pages = "015023",
    year = "2019"
}

@article{Bauer:2019vqk,
    author = "Bauer, Martin and Brandt, Oleg and Lee, Lawrence and Ohm, Christian",
    title = "{ANUBIS: Proposal to search for long-lived neutral particles in CERN service shafts}",
    eprint = "1909.13022",
    archivePrefix = "arXiv",
    primaryClass = "physics.ins-det",
    month = "9",
    year = "2019"
}

@article{Alpigiani:2020tva,
    author = "Alpigiani, Cristiano and others",
    collaboration = "MATHUSLA",
    title = "{An Update to the Letter of Intent for MATHUSLA: Search for Long-Lived Particles at the HL-LHC}",
    eprint = "2009.01693",
    archivePrefix = "arXiv",
    primaryClass = "physics.ins-det",
    reportNumber = "CERN-LHCC-2020-014, LHCC-I-031-ADD-1",
    month = "9",
    year = "2020"
}

@article{Minkowski:1977sc,
    author = "Minkowski, Peter",
    title = "{$\mu \to e\gamma$ at a Rate of One Out of $10^{9}$ Muon Decays?}",
    reportNumber = "Print-77-0182 (BERN)",
    doi = "10.1016/0370-2693(77)90435-X",
    journal = "Phys. Lett. B",
    volume = "67",
    pages = "421--428",
    year = "1977"
}

@article{GellMann:1980vs,
    author = "Gell-Mann, Murray and Ramond, Pierre and Slansky, Richard",
    title = "{Complex Spinors and Unified Theories}",
    eprint = "1306.4669",
    archivePrefix = "arXiv",
    primaryClass = "hep-th",
    reportNumber = "PRINT-80-0576",
    journal = "Conf. Proc. C",
    volume = "790927",
    pages = "315--321",
    year = "1979"
}

@article{Mohapatra:1979ia,
    author = "Mohapatra, Rabindra N. and Senjanovic, Goran",
    title = "{Neutrino Mass and Spontaneous Parity Nonconservation}",
    reportNumber = "MDDP-TR-80-060, MDDP-PP-80-105, CCNY-HEP-79-10",
    doi = "10.1103/PhysRevLett.44.912",
    journal = "Phys. Rev. Lett.",
    volume = "44",
    pages = "912",
    year = "1980"
}

@article{Yanagida:1980xy,
    author = "Yanagida, Tsutomu",
    title = "{Horizontal Symmetry and Masses of Neutrinos}",
    reportNumber = "TU-80-208",
    doi = "10.1143/PTP.64.1103",
    journal = "Prog. Theor. Phys.",
    volume = "64",
    pages = "1103",
    year = "1980"
}

@article{Schechter:1980gr,
    author = "Schechter, J. and Valle, J.W.F.",
    title = "{Neutrino Masses in SU(2) x U(1) Theories}",
    reportNumber = "SU-4217-167, COO-3533-167",
    doi = "10.1103/PhysRevD.22.2227",
    journal = "Phys. Rev. D",
    volume = "22",
    pages = "2227",
    year = "1980"
}

@article{Schechter:1981cv,
    author = "Schechter, J. and Valle, J.W.F.",
    title = "{Neutrino Decay and Spontaneous Violation of Lepton Number}",
    reportNumber = "SU-4217-203, COO-3533-203",
    doi = "10.1103/PhysRevD.25.774",
    journal = "Phys. Rev. D",
    volume = "25",
    pages = "774",
    year = "1982"
}

@article{Fukugita:1986hr,
    author = "Fukugita, M. and Yanagida, T.",
    title = "{Baryogenesis Without Grand Unification}",
    reportNumber = "RIFP-641",
    doi = "10.1016/0370-2693(86)91126-3",
    journal = "Phys. Lett. B",
    volume = "174",
    pages = "45--47",
    year = "1986"
}

@article{Dodelson:1993je,
    author = "Dodelson, Scott and Widrow, Lawrence M.",
    title = "{Sterile-neutrinos as dark matter}",
    eprint = "hep-ph/9303287",
    archivePrefix = "arXiv",
    reportNumber = "FERMILAB-PUB-93-057-A",
    doi = "10.1103/PhysRevLett.72.17",
    journal = "Phys. Rev. Lett.",
    volume = "72",
    pages = "17--20",
    year = "1994"
}

@article{Blondel:2014bra,
    author = "Blondel, Alain and Graverini, E. and Serra, N. and Shaposhnikov, M.",
    editor = "Aguilar-Ben\'\i{}tez, M and Fuster, J and Mart\'\i{}-Garc\'\i{}a, S and Santamar\'\i{}a, A",
    collaboration = "FCC-ee study Team",
    title = "{Search for Heavy Right Handed Neutrinos at the FCC-ee}",
    eprint = "1411.5230",
    archivePrefix = "arXiv",
    primaryClass = "hep-ex",
    doi = "10.1016/j.nuclphysbps.2015.09.304",
    journal = "Nucl. Part. Phys. Proc.",
    volume = "273-275",
    pages = "1883--1890",
    year = "2016"
}

@article{Drewes:2013gca,
    author = "Drewes, Marco",
    title = "{The Phenomenology of Right Handed Neutrinos}",
    eprint = "1303.6912",
    archivePrefix = "arXiv",
    primaryClass = "hep-ph",
    reportNumber = "TUM-HEP-881-13",
    doi = "10.1142/S0218301313300191",
    journal = "Int. J. Mod. Phys. E",
    volume = "22",
    pages = "1330019",
    year = "2013"
}

@article{Garbrecht:2018mrp,
    author = {Garbrecht, Bj\"orn},
    title = "{Why is there more matter than antimatter? Calculational methods for leptogenesis and electroweak baryogenesis}",
    eprint = "1812.02651",
    archivePrefix = "arXiv",
    primaryClass = "hep-ph",
    reportNumber = "TUM-HEP-1177-18",
    doi = "10.1016/j.ppnp.2019.103727",
    journal = "Prog. Part. Nucl. Phys.",
    volume = "110",
    pages = "103727",
    year = "2020"
}

@article{Bodeker:2020ghk,
    author = "Bodeker, Dietrich and Buchmuller, Wilfried",
    title = "{Baryogenesis from the weak scale to the GUT scale}",
    eprint = "2009.07294",
    archivePrefix = "arXiv",
    primaryClass = "hep-ph",
    reportNumber = "DESY-20-141",
    month = "9",
    year = "2020"
}

@article{Chun:2017spz,
    author = "Chun, E.J. and others",
    title = "{Probing Leptogenesis}",
    eprint = "1711.02865",
    archivePrefix = "arXiv",
    primaryClass = "hep-ph",
    doi = "10.1142/S0217751X18420058",
    journal = "Int. J. Mod. Phys. A",
    volume = "33",
    number = "05n06",
    pages = "1842005",
    year = "2018"
}

@article{Adhikari:2016bei,
    author = "Drewes, M. and others",
    title = "{A White Paper on keV Sterile Neutrino Dark Matter}",
    eprint = "1602.04816",
    archivePrefix = "arXiv",
    primaryClass = "hep-ph",
    reportNumber = "FERMILAB-PUB-16-068-T",
    doi = "10.1088/1475-7516/2017/01/025",
    journal = "JCAP",
    volume = "01",
    pages = "025",
    year = "2017"
}

@article{Boyarsky:2018tvu,
    author = "Boyarsky, A. and Drewes, M. and Lasserre, T. and Mertens, S. and Ruchayskiy, O.",
    title = "{Sterile neutrino Dark Matter}",
    eprint = "1807.07938",
    archivePrefix = "arXiv",
    primaryClass = "hep-ph",
    doi = "10.1016/j.ppnp.2018.07.004",
    journal = "Prog. Part. Nucl. Phys.",
    volume = "104",
    pages = "1--45",
    year = "2019"
}

@article{Atre:2009rg,
    author = "Atre, Anupama and Han, Tao and Pascoli, Silvia and Zhang, Bin",
    title = "{The Search for Heavy Majorana Neutrinos}",
    eprint = "0901.3589",
    archivePrefix = "arXiv",
    primaryClass = "hep-ph",
    reportNumber = "FERMILAB-PUB-08-086-T, NSF-KITP-08-54, MADPH-06-1466, DCPT-07-198, IPPP-07-99",
    doi = "10.1088/1126-6708/2009/05/030",
    journal = "JHEP",
    volume = "05",
    pages = "030",
    year = "2009"
}

@article{Deppisch:2015qwa,
    author = "Deppisch, Frank F. and Bhupal Dev, P.S. and Pilaftsis, Apostolos",
    title = "{Neutrinos and Collider Physics}",
    eprint = "1502.06541",
    archivePrefix = "arXiv",
    primaryClass = "hep-ph",
    reportNumber = "MAN-HEP-2014-15",
    doi = "10.1088/1367-2630/17/7/075019",
    journal = "New J. Phys.",
    volume = "17",
    number = "7",
    pages = "075019",
    year = "2015"
}

@article{Cai:2017mow,
    author = "Cai, Yi and Han, Tao and Li, Tong and Ruiz, Richard",
    title = "{Lepton Number Violation: Seesaw Models and Their Collider Tests}",
    eprint = "1711.02180",
    archivePrefix = "arXiv",
    primaryClass = "hep-ph",
    reportNumber = "PITT-PACC-1712, IPPP-17-74, COEPP-MN-17-17",
    doi = "10.3389/fphy.2018.00040",
    journal = "Front. in Phys.",
    volume = "6",
    pages = "40",
    year = "2018"
}

@article{Antusch:2016ejd,
    author = "Antusch, Stefan and Cazzato, Eros and Fischer, Oliver",
    title = "{Sterile neutrino searches at future $e^-e^+$, $pp$, and $e^-p$ colliders}",
    eprint = "1612.02728",
    archivePrefix = "arXiv",
    primaryClass = "hep-ph",
    doi = "10.1142/S0217751X17500786",
    journal = "Int. J. Mod. Phys. A",
    volume = "32",
    number = "14",
    pages = "1750078",
    year = "2017"
}

@article{Johnson:1997cj,
    author = "Johnson, Loretta M. and McKay, Douglas W. and Bolton, Tim",
    title = "{Extending sensitivity for low mass neutral heavy lepton searches}",
    eprint = "hep-ph/9703333",
    archivePrefix = "arXiv",
    doi = "10.1103/PhysRevD.56.2970",
    journal = "Phys. Rev. D",
    volume = "56",
    pages = "2970--2981",
    year = "1997"
}

@article{Gorbunov:2007ak,
    author = "Gorbunov, Dmitry and Shaposhnikov, Mikhail",
    title = "{How to find neutral leptons of the $\nu$MSM?}",
    eprint = "0705.1729",
    archivePrefix = "arXiv",
    primaryClass = "hep-ph",
    doi = "10.1088/1126-6708/2007/10/015",
    journal = "JHEP",
    volume = "10",
    pages = "015",
    year = "2007",
    related = "Gorbunov:2007erratum",
    relatedtype= "erratum",
}

@article{Canetti:2012kh,
    author = "Canetti, Laurent and Drewes, Marco and Frossard, Tibor and Shaposhnikov, Mikhail",
    title = "{Dark Matter, Baryogenesis and Neutrino Oscillations from Right Handed Neutrinos}",
    eprint = "1208.4607",
    archivePrefix = "arXiv",
    primaryClass = "hep-ph",
    reportNumber = "TTK-12-05, TUM-HEP-852-12, CAS-KITPC-ITP-368",
    doi = "10.1103/PhysRevD.87.093006",
    journal = "Phys. Rev. D",
    volume = "87",
    pages = "093006",
    year = "2013"
}

@article{Bondarenko:2018ptm,
    author = "Bondarenko, Kyrylo and Boyarsky, Alexey and Gorbunov, Dmitry and Ruchayskiy, Oleg",
    title = "{Phenomenology of GeV-scale Heavy Neutral Leptons}",
    eprint = "1805.08567",
    archivePrefix = "arXiv",
    primaryClass = "hep-ph",
    doi = "10.1007/JHEP11(2018)032",
    journal = "JHEP",
    volume = "11",
    pages = "032",
    year = "2018"
}

@article{Ballett:2019bgd,
    author = "Ballett, Peter and Boschi, Tommaso and Pascoli, Silvia",
    title = "{Heavy Neutral Leptons from low-scale seesaws at the DUNE Near Detector}",
    eprint = "1905.00284",
    archivePrefix = "arXiv",
    primaryClass = "hep-ph",
    reportNumber = "IPPP/18/76",
    doi = "10.1007/JHEP03(2020)111",
    journal = "JHEP",
    volume = "03",
    pages = "111",
    year = "2020"
}

@article{Coloma:2020lgy,
    author = "Coloma, Pilar and Fern\'andez-Mart\'\i{}nez, Enrique and Gonz\'alez-L\'opez, Manuel and Hern\'andez-Garc\'\i{}a, Josu and Pavlovic, Zarko",
    title = "{GeV-scale neutrinos: interactions with mesons and DUNE sensitivity}",
    eprint = "2007.03701",
    archivePrefix = "arXiv",
    primaryClass = "hep-ph",
    reportNumber = "FERMILAB-PUB-20-269-ND",
    month = "7",
    year = "2020"
}

@article{Pascoli:2018heg,
    author = "Pascoli, Silvia and Ruiz, Richard and Weiland, Cedric",
    title = "{Heavy neutrinos with dynamic jet vetoes: multilepton searches at $ \sqrt{s}=14$, $27$, and $100$ TeV}",
    eprint = "1812.08750",
    archivePrefix = "arXiv",
    primaryClass = "hep-ph",
    reportNumber = "CP3-18-77, IPPP/18/111, PITT-PACC-1821, VBSCAN-PUB-10-18",
    doi = "10.1007/JHEP06(2019)049",
    journal = "JHEP",
    volume = "06",
    pages = "049",
    year = "2019"
}

@article{deVries:2020qns,
    author = {de Vries, Jordy and Dreiner, Herbert K. and G\"unther, Julian Y. and Wang, Zeren Simon and Zhou, Guanghui},
    title = "{Long-lived Sterile Neutrinos at the LHC in Effective Field Theory}",
    eprint = "2010.07305",
    archivePrefix = "arXiv",
    primaryClass = "hep-ph",
    reportNumber = "APCTP Pre2020-027, BONN-TH-2020-10, RBRC-1328",
    month = "10",
    year = "2020"
}

@article{Drewes:2019vjy,
    author = "Drewes, Marco and Giammanco, Andrea and Hajer, Jan and Lucente, Michele",
    title = "{New long-lived particle searches in heavy-ion collisions at the LHC}",
    eprint = "1905.09828",
    archivePrefix = "arXiv",
    primaryClass = "hep-ph",
    reportNumber = "CP3-19-26",
    doi = "10.1103/PhysRevD.101.055002",
    journal = "Phys. Rev. D",
    volume = "101",
    number = "5",
    pages = "055002",
    year = "2020"
}

@article{Sjostrand:2014zea,
    author = {Sj\"ostrand, Torbj\"orn and Ask, Stefan and Christiansen, Jesper R. and Corke, Richard and Desai, Nishita and Ilten, Philip and Mrenna, Stephen and Prestel, Stefan and Rasmussen, Christine O. and Skands, Peter Z.},
    title = "{An introduction to PYTHIA 8.2}",
    eprint = "1410.3012",
    archivePrefix = "arXiv",
    primaryClass = "hep-ph",
    reportNumber = "LU-TP-14-36, MCNET-14-22, CERN-PH-TH-2014-190, FERMILAB-PUB-14-316-CD, DESY-14-178, SLAC-PUB-16122",
    doi = "10.1016/j.cpc.2015.01.024",
    journal = "Comput. Phys. Commun.",
    volume = "191",
    pages = "159--177",
    year = "2015"
}

@article{Aaboud:2017iio,
    author = "Aaboud, Morad and others",
    collaboration = "ATLAS",
    title = "{Search for long-lived, massive particles in events with displaced vertices and missing transverse momentum in $\sqrt{s}$ = 13 TeV $pp$ collisions with the ATLAS detector}",
    eprint = "1710.04901",
    archivePrefix = "arXiv",
    primaryClass = "hep-ex",
    reportNumber = "CERN-EP-2017-202",
    doi = "10.1103/PhysRevD.97.052012",
    journal = "Phys. Rev. D",
    volume = "97",
    number = "5",
    pages = "052012",
    year = "2018"
}

@techreport{ATLAS:2019wqx,
    collaboration = "ATLAS",
    title = "{Performance of vertex reconstruction algorithms for detection of new long-lived particle decays within the ATLAS inner detector}",
    number = "ATL-PHYS-PUB-2019-013",
    year = "2019"
}

@article{Buskulic:1994wz,
    author = "Buskulic, D. and others",
    collaboration = "ALEPH",
    title = "{Performance of the ALEPH detector at LEP}",
    reportNumber = "CERN-PPE-94-170, FSU-SCRI-95-70",
    doi = "10.1016/0168-9002(95)00138-7",
    journal = "Nucl. Instrum. Meth. A",
    volume = "360",
    pages = "481--506",
    year = "1995"
}

@techreport{LHCbCollaboration:2014tuj,
    collaboration = "LHCb",
    title = "{LHCb Tracker Upgrade Technical Design Report}",
    number = "CERN-LHCC-2014-001, LHCB-TDR-015",
    month = "2",
    year = "2014"
}

\glsunset{THUNDERDOME}

\end{document}